# Bipolar High Power Impulse Magnetron Sputtering for energetic ion bombardment during TiN thin film growth without the use of a substrate bias

Rommel Paulo B. Viloan[a†], Jiabin Gu[b†], Robert Boyd[a], Julien Keraudy[a*], Liuhe Li[b], Ulf Helmersson[a]

[a] *Department of Physics, Linköping University, Linköping, SE-581 83, Sweden*

[b] *School of Mechanical Engineering and Automation, Beihang University, Beijing, 100191, PR China*

**Abstract**

The effect of applying a positive voltage pulse ($U_{rev}$ = 10 - 150 V) directly after the negative high power impulse magnetron sputtering (HiPIMS) pulse (bipolar HiPIMS) is investigated for the reactive sputter deposition of TiN thin films. Energy-resolved mass spectroscopy analyses are performed to gain insight in the effect on the ion energy distribution function of the various ions. It is demonstrated that the energy of a large fraction of the ions can be tuned by a reverse target potential and gain energy corresponding to the applied $U_{rev}$. Microscopy observations and x-ray reflectometry reveal densification of the films which results in an increase in the film hardness from 23.9 to 34 GPa as well as an increase in compressive film stress from 2.1 GPa to 4.7 GPa when comparing conventional HiPIMS with bipolar HiPIMS ($U_{rev}$ = 150 V).

**Introduction**

High power impulse magnetron sputtering (HiPIMS) is a physical vapor deposition technique that has attracted significant attention during the past years. [1–4] This is due to its relatively simple implementation to strengthen conventional magnetron sputtering processes. When compared with direct current magnetron sputtering, HiPIMS is known to produce a significant amount of ions of the sputtered material with a slightly broadened ion energy distribution function (IEDF) which both has implications to the controllability of the process and quality of the deposited films. [5–8] These process improvements are a result of the high power density, several kW/cm$^2$, that are applied to the cathode during the pulses of the HiPIMS discharge.

---

[†] *These authors contributed equally to the work.*

[*] *Current affiliation: Oerlikon Surface Solutions AG, Oerlikon Balzers, Iramali 18, LI-9496, Balzers, Liechtenstein*

Despite a broad ion energy distribution, the HiPIMS discharge is still dominated by thermalized ions. Therefore, the control of the energy of the incoming ion flux to the growing film is generally achieved by applying a substrate bias. This, however, becomes challenging when depositing insulating films or using non-conducting substrates and even impossible when considering grounded substrates that can occur in some industrial deposition systems. Recently, the introduction of bipolar HiPIMS wherein a reversed (positive) pulse is applied to the target following the negative pulse has promised great potential to solve these challenges. In this mode of operation a portion of the IEDF can be shifted with an energy proportional to the magnitude of the applied reversed potential, $U_{rev}$. This is a consequence of the fact that a region of the plasma, near the cathode, achieve an increased plasma potential with a value close to $U_{rev}$ resulting in acceleration of ions leaving this near-target-region. [9,10]

By using HiPIMS in bipolar mode, film properties can be improved due to energetic bombardment of ions. Velicu *et al.* [9] have shown that Cu film adhesion, density and microstructure can be tailored depending on the magnitude of $U_{rev}$. Using the same film material, Wu *et al.* [11] demonstrated an increase in deposition rate arguing that the return effect in HiPIMS is reduced due to the application of a positive pulse and the films exhibit reduced tensile stress. Britun *et al.* [12] were able to observe a shift in the texture of Ti films when increasing the magnitude of $U_{rev}$.

Earlier we reported on the effect of the application of bipolar HiPIMS on the IEDF of the ions during sputtering of Ti in pure Ar. [10] There we concluded that the application of a positive pulse potential following the negative HiPIMS pulse effectively tailors the IEDF of the ions. In this work, we apply the same technique and explore the effect on reactively deposited TiN films. It is shown that the IEDF of the ions can be effectively tuned and the properties of the TiN film are enhanced.

**Experimental Details**

The experiments are performed in a cylindrical high vacuum system (with a base pressure of $1.3\times10^{-4}$ Pa, a radius of 22 cm and a height of 30 cm) equipped with a sputtering magnetron cathode (TORUS Circular, Kurt J. Lesker Company). A Ti target (50 mm in diameter, 99.9% purity) is mounted on the cathode. During sputtering an Ar flow (99.997% purity) of 50 sccm and a $N_2$ flow (99.995% purity) of 0.35 sccm is used. The total pressure is maintained at 0.66 Pa using a throttle valve. The cathode is connected to a pulsing unit which is fed with negative and positive pulse potentials as programmed by a synchronization unit (a HiPSTER prototype from Ionautics AB). The negative pulse (the conventional HiPIMS pulse) has a length of 30 μs and is operated at a repetition frequency of 700 Hz. Each individual HiPIMS pulse is immediately followed by a 200 μs long positive pulse when operated in bipolar mode. The voltage and current waveforms are recorded through a 1:100 voltage divider and a current clamp (Chauvin Arnaux C160) connected to a digital oscilloscope (Tektronix TDS 2004C).

A PSM003 ion mass spectrometer (Hiden Analytical Ltd) capable of measuring ion energies up to 100 eV is used for measuring the IEDF of $Ar^+$ (36 amu), $Ar^{2+}$ (40 amu), $Ti^+$ (48 amu), $Ti^{2+}$ (48 amu), $N^+$ (14 amu) and $N_2^+$ (28 amu) ion species. The sampling orifice, which is aligned with the center of the Ti target at a distance of ~8 cm, has an opening of 300 μm in diameter and is grounded during the measurements. The energy step size is set to 0.1 eV while the acquisition per data point is set to 200 ms corresponding to 140 pulses averaged per data point. During the acquisition of the IEDF presented in this work the reverse voltage ($U_{rev}$) was set to 0, 25, 50 and 75 V.

For thin film deposition substrates of Si (100) with dimensions of 1 × 1 cm are used. Prior to deposition the substrates are cleaned in an ultrasonic bath of acetone followed by a bath of isopropyl alcohol for a total time of 10 minutes. The samples are blown dry using $N_2$ and mounted in a rotatable sample holder which exposes one sample at a time at a distance of ~8 cm away from the target. The reverse voltage ($U_{rev}$) was set to 0, 10, 50, 100 and 150 V and the substrates were kept at floating potential during the depositions presented in this work.

X-ray diffraction (XRD) is done using a PANalytical X'Pert PRO in Bragg Brentano (Θ-2Θ) geometry for texture characterization of the films while a PANalytical Empyrean system with a four-crystal Ge (220) monochromator and a two-bounce triple axis Ge (220) analyzer is used for Ω rocking curve measurements from which the curvature of a single crystal substrate can be derived [13,14]. The Stoney equation for anisotropic single crystal Si (100) substrate is then applied to calculate the film stress from the substrate curvature [15]. The X-ray source in both diffractometers emit a $CuK\alpha_1$ radiation (λ = 1.540597 Å) operated at 45 kV and 40 mA and all measurements are calibrated to the Si (100) peak at 69.13°. The film density is measured by X-ray reflectometry (XRR) using the Empyrean system but the incident optics is changed to a two-bounce Ge (220) hybrid mirror. The fitting of the XRR data is done using a PANalytical X'Pert Reflectivity software and the error of the density measurement is taken from the error analysis of the fitting procedure.

Nanoindentation analyses is done using a Hysitron TI-950 Triboindenter equipped with a Berkovich diamond probe. A constant load of 0.6 mN, for which the indentation depth did not exceed 10% of the film thickness, is used to execute a minimum of 15 indents in each sample.

The microstructure was investigated by transmission electron microscopy (TEM) (FEI Tecnai G2 operated at 200 kV). Prior to analysis, cross-sectional specimens are prepared by a two-step procedure consisting of mechanical polishing followed by $Ar^+$ milling at shallow incidence angle of 4° from the sample surface, with 5 keV ions initially then the ion energy was decreased to 2.5 and 1 keV. High angle angular dark field (HAADF) combined with scanning-TEM (STEM) analysis is obtained using a camera length of 140 mm. The film deposition rate is calculated using observed film thicknesses from scanning electron microscopy (SEM) (Leo 1550 Gemini) images of film cross-sections. Each film is imaged in three different places across the sample length.

**Results**

The discharge voltage ($U_D$) and discharge current ($I_D$) waveforms recorded are shown in Fig. 1 for applied $U_{rev}$ values of 0, 50 and 150 V. A constant power ($P_{avg}$) of 80 W were used for which a $U_{rev}$ of 0 and 50 V required a $U_D$ = -450 V and for a $U_{rev}$ of 150 V required -464 V. The $I_D$ waveform changes slightly when $U_{rev}$ is varied with an increased delay in current rise with an increase in $U_{rev}$ from 4.9 µs to 5.3 µs and finally to 9.2 µs for the different $U_{rev}$ values, respectively. Accordingly, the $I_{D, max}$ also increases with $U_{rev}$ as a consequence of the constant $P_{avg}$ and constant pulse length applied.

Fig. 2 shows the time-averaged IEDF of $Ar^+$, $Ti^+$ and $N^+$ acquired for $U_{rev}$ set at 0, 25, 50 and 75 V. The $Ar^+$, $Ti^+$ and $N^+$ IEDFs show a sharp low energy peak (LE) at ~1 eV for a conventional HiPIMS discharge ($U_{rev}$ = 0 V). This is followed by a shoulder which peaks at ~10 eV for $Ti^+$ and $N^+$ while a sharp drop in the energy distribution is observed for $Ar^+$. An extensive high energy tail (HE) is observed for $Ti^+$ and $N^+$ which is similar to what has been observed with a HiPIMS discharge of Ti in $Ar/N_2$ environments. [16,17] When a $U_{rev}$ is applied, the LE, shoulder and HE intensities decrease while a narrow high intensity peak at an energy approximately equal to the applied $U_{rev}$ appears with the same energy profile as the IEDF of the conventional HiPIMS pulse. This observed upshifting of the ion energy distribution is similar to what was observed by Keraudy *et al*. [10] for a bipolar HiPIMS discharge of Ti sputtering in pure Ar.

Fig 3 shows the Ɵ-2Ɵ XRD scans of the deposited TiN films. The scans where done over the range from 30 to 80º, but only the details over the most intense TiN peaks are shown in the figure. Peaks corresponding to (220) and (222) are also detected but with lower intensities than the (111) or (200) peaks shown here. The relative peak intensities, the film texture, remains close to constant for the range of growth parameters investigated. It is clear from the figure that there is a shift to lower 2Θ values as $U_{rev}$ is increased. With respect to the unstrained peaks of (111) and (200) at 36.66º and 42.59º, respectively, the film deposited with $U_{rev} = 0$ is shifted to lower values by about 0.1º and 0.2º. The (111) and (200) peaks are shifted to even lower values by as much as ~0.4º for $U_{rev} = 150$ V. The full width at half maximum of both peaks increases with an increase in $U_{rev}$ although the increase is only about 10 %.

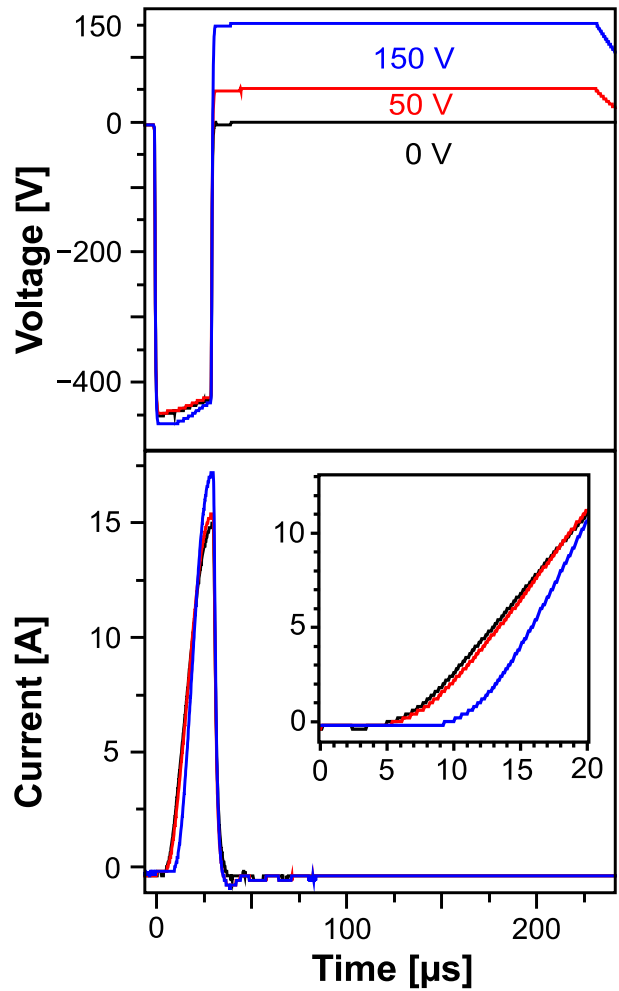

Figure 1. Voltage and current waveforms recorded for 0, 50 and 100 V reverse voltages. The inset in the current waveform shows the initial evolution of the current waveform.

The film stress, hardness and density are plotted in Fig. 4. The compressive stress of the thin films increases as $U_{rev}$ is increased. A compressive stress of ~2.1 GPa is measured for the film deposited using $U_{rev} = 0$. Applying $U_{rev} = 10$ V has a significant effect on the film's compressive stress which changed to ~3.1 GPa. The compressive stress continually increases as $U_{rev}$ increases and reaches a maximum of ~4.7 GPa for $U_{rev} = 150$ V.

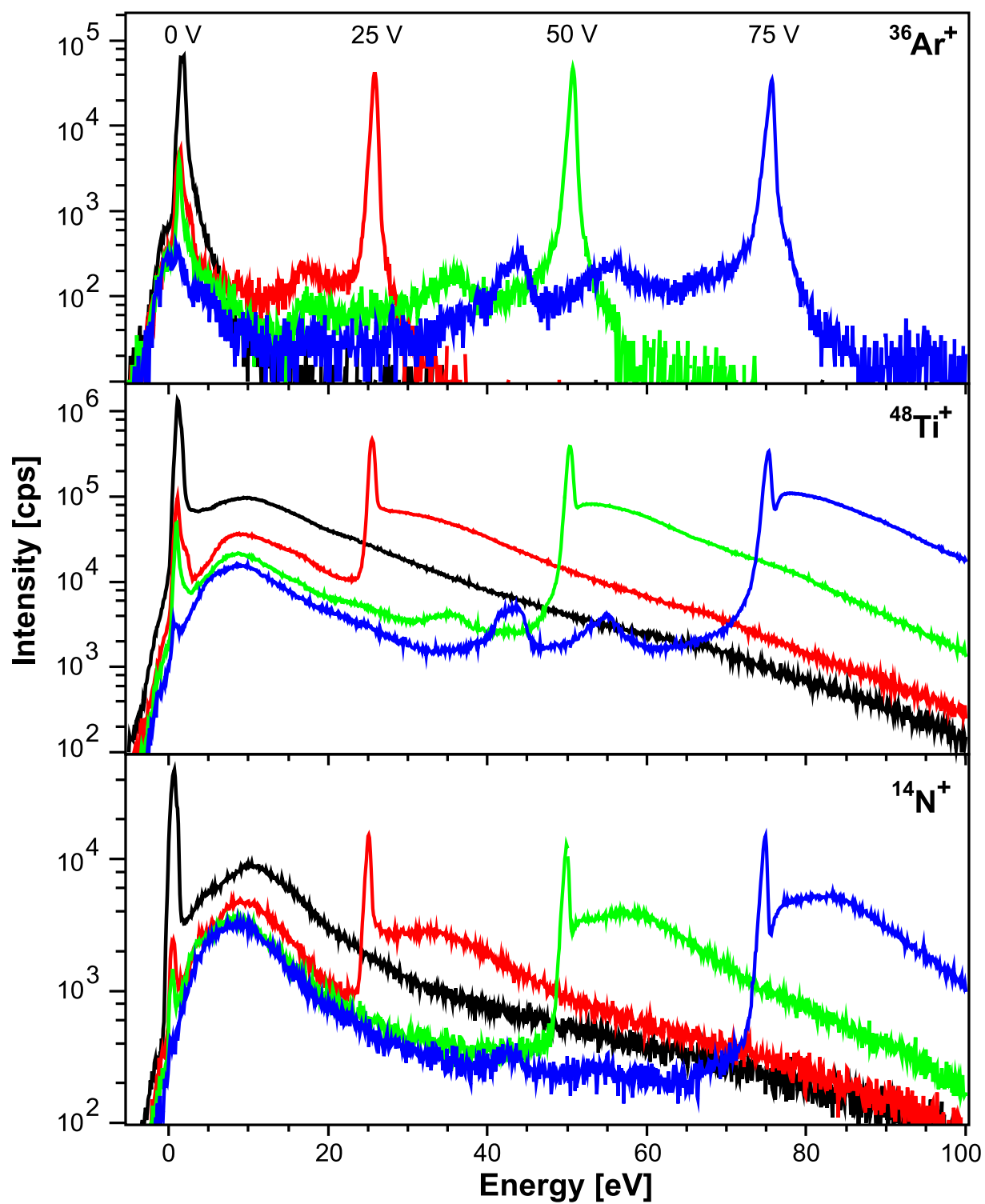

Figure 2. Time-averaged IEDF measured at the substrate position during bipolar HiPIMS sputtering of Ti in Ar/$N_2$. The voltage values given in the figure indicates the curves collected for the different voltage level of the reversed second pulse.

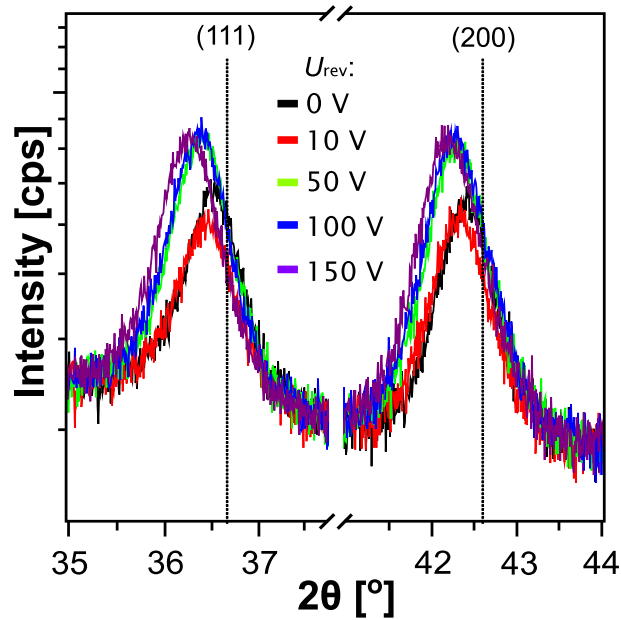

Figure 3. XRD Θ-2Θ of TiN films grown using different applied bipolar pulse potential, $U_{rev}$

The same trend is observed for the measured hardness. As $U_{rev}$ is increased the hardness increases correspondingly. There is a noticeable increase in hardness from ~23.9 to ~29.32 GPa when applying only 10 V. The hardness continually increases to a maximum of ~34 GPa for $U_{rev}$ = 150 V.

XRR measurements reveal a clear shift to higher critical angle (not shown) when $U_{rev}$ is increased which means that the film density increases with the application of bipolar HiPIMS. The density is calculated from these measurements and gives a film density of ~5.1 g/cm$^3$ for $U_{rev}$ = 0 V, ~5.2 g/cm$^3$ for $U_{rev}$ = 10 V, and a maximum of ~5.3 g/cm$^3$ for $U_{rev}$ =150 V. The density errors given in Fig. 4 are absolute errors - relative errors are smaller.

TEM analysis of the selected TiN films, presented in Fig. 5, shows columnar structures. In the case of the film deposited with $U_{rev}$ = 150 V, there is small displacement between the film and the silicon substrate, indicating potential delamination. Representative HAADF-STEM micrographs from the films are also shown in Fig. 5. In such figures contrast is primarily determined by the thickness of the sample, assuming a homogenous composition [18]. The grain boundaries are clearly distinguishable in the case of $U_{rev}$ = 0 and become significantly less so with increasing $U_{rev}$. The most likely explanation of the observed behavior is that the grain boundaries are underdense for films deposited with $U_{rev}$ = 0 V and become increasingly denser with $U_{rev}$.

The deposition rates are shown in Fig. 6 and indicate a small (~15 %) decrease in rate as $U_{rev}$ increase from 0 to 150 V. The main part of this decrease occurs already when applying $U_{rev}$ = 10 V where after the changes are within the estimated error of the measurements.

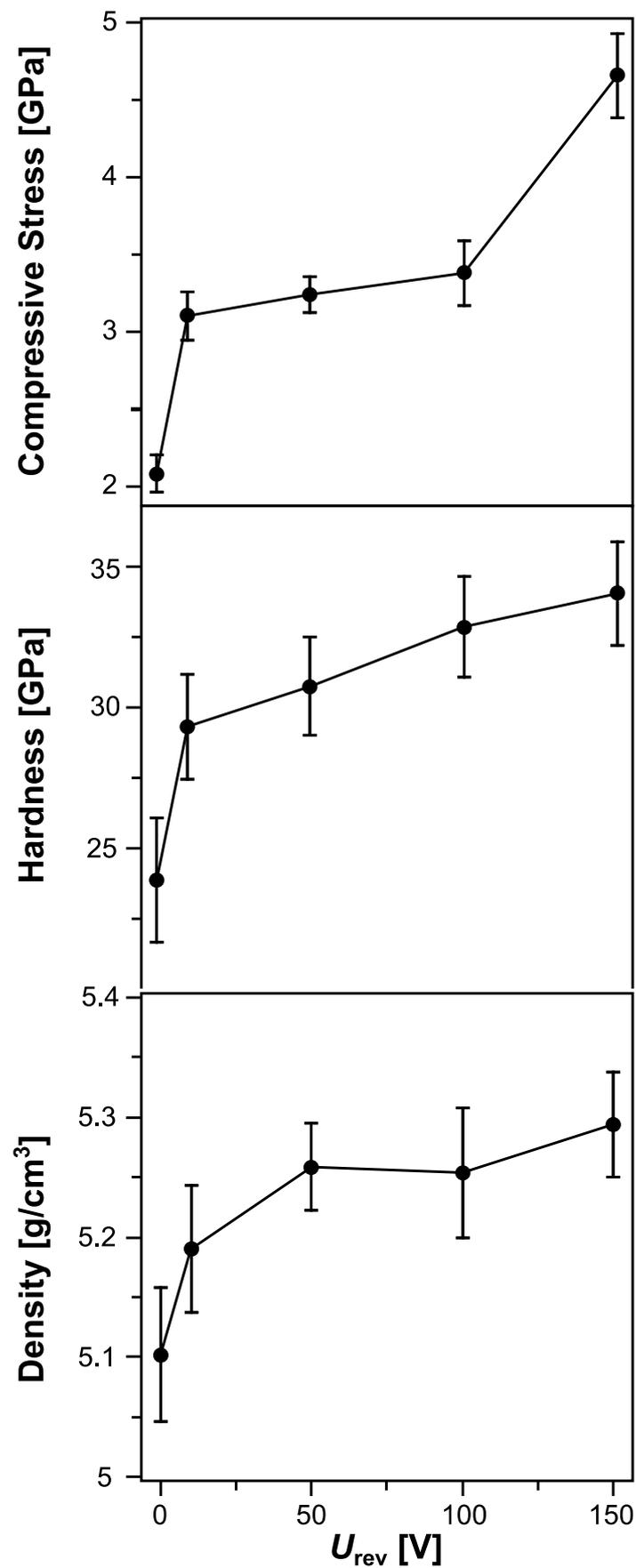

Figure 4. The measured stress, hardness and density of the deposited TiN thin films as a function of the reversed pulse potential, $U_{rev}$.

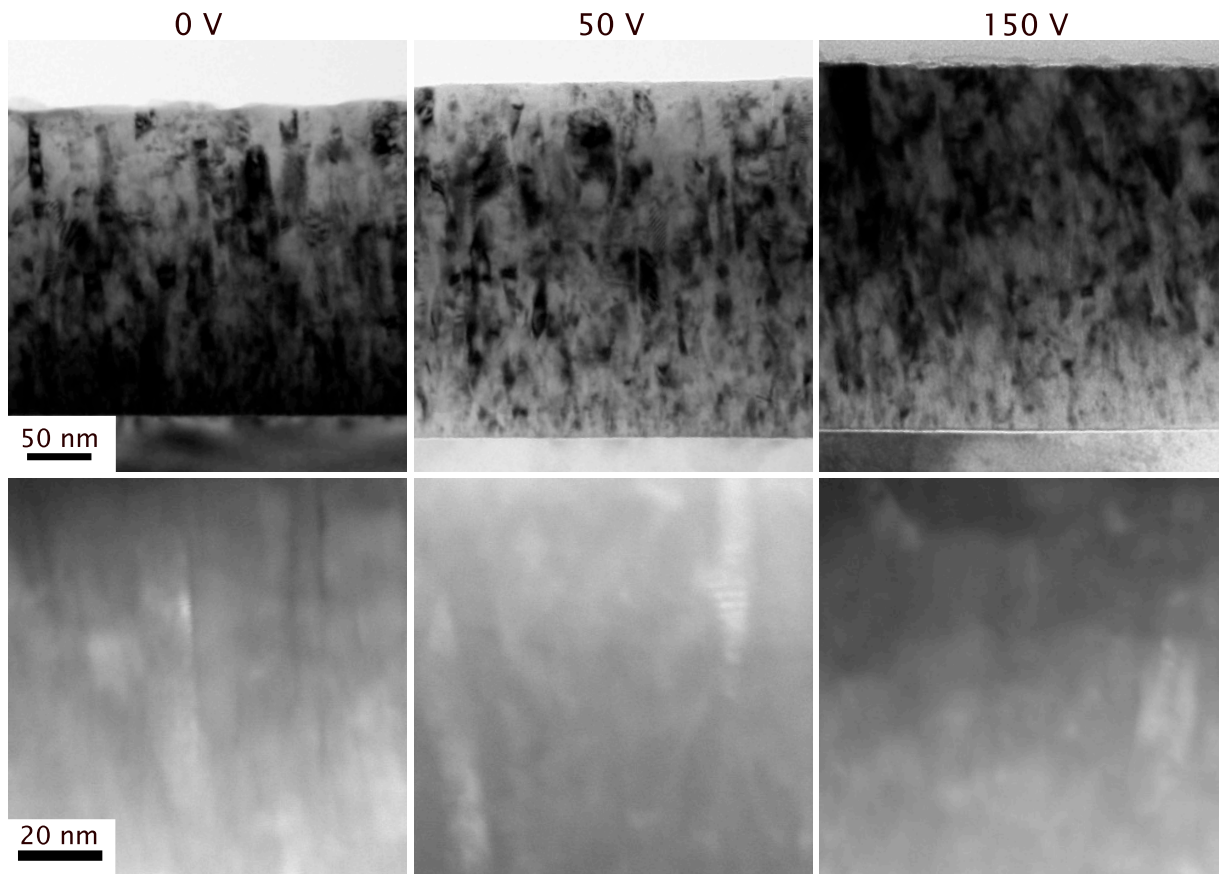

Figure 5. (Top) Bright field TEM and (bottom) HAADF-STEM image from TiN films deposited with $U_{rev}$ = 0 V (a), 50 V (b) and 150 V (c).

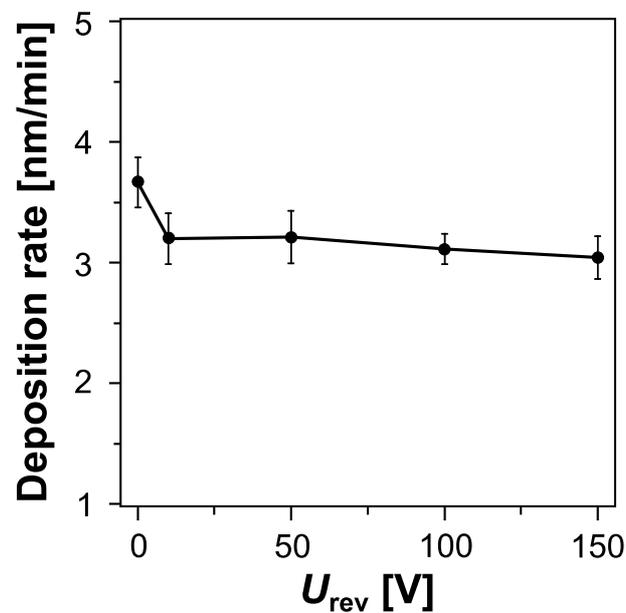

Figure 6. Deposition rate measured from thickness measurement of the TiN films in SEM normalized by the individual deposition times for each parrameter.

**Discussion**

It is clear from the results presented in Fig. 1 that the ignition of the HiPIMS pulse is affected by the reversed pulse resulting in a delay in ignition that increases with increasing $U_{rev}$. Poolcharuansin and Bradley have earlier shown that during the off-time period the electron density decays by two orders of magnitude. [19] They conclude that the decay is fast for hot electrons (<100 μs) while cold electrons maintain a relatively high plasma density even 10 ms after the HiPIMS-pulse. For HiPIMS repetition rate used in the present study (700 Hz, 1.4 ms), a significant amount of charge carriers should remain from one pulse to the other. The application of a positive target potential after the negative pulse will likely increase the drainage of charge carriers from the plasma in-between the HiPIMS-pulses. Thus, a starting point with less charge carriers will be the result, that delay the current rise of the succeeding pulse, a delay that increases as $U_{rev}$ is increased.

The IEDFs for $Ti^+$ and for $N^+$ of the conventional HiPIMS discharge, shown in Fig. 2 (0 V), have similar appearance, with a narrow low energy peak followed by a shoulder and a high-energy tail. This differs significantly from the appearance of the $Ar^+$ IEDF, which has no shoulder and only a shorter high energy tail. $N_2^+$ ions (not shown) show an IEDF similar to the one for $Ar^+$. The IEDFs for $Ti^{2+}$ and $Ar^{2+}$ (not shown) follow the same trend as their singly ionized counterpart, but at a lower intensity. The difference in IEDFs of these species originate in how the species are created. Ti and N atoms are generated through sputtering of the target, and will attain a Thompson energy distribution which is significantly hotter (and with a longer high energy tail) than the thermal process gas consisting of Ar and $N_2$. [20] The other significant portion of the IEDFs of $Ti^+$ and $N^+$ is the shoulder which may be explained by the presence of self-organizing ionization zones – spokes. [20,21] The process gas species are not significantly affected by these spokes since the region close to the sputtering target to a large extent is diluted from the gases during the most intense part of the HiPIMS pulse.

The IEDFs for the bipolar HiPIMS seen in Fig. 2, (25, 50 and 75 V), are characterized by a low energy peak, as well as a shoulder, that decreases with the applied $U_{rev}$. Added to this is a high energy peak, including the shoulder, that appears at an energy corresponding to the applied $U_{rev}$ plus the energy of the original low energy peak and shoulder. It is clear that a growing portion of the original ion-population is accelerated over the full potential as $U_{rev}$ increases. Such acceleration of ions by the reverse voltage have been observed in several studies. [9,10,22,23] A model of this phenomena was published earlier [10] which suggests that the fraction of the plasma volume in good contact with the sputtering target and no contact with anode (that is the plasma volume containing electrons that can reach the target surface without crossing any magnetic field lines, but need to cross field lines to reach the anode) quickly attain a plasma potential slightly higher than $U_{rev}$, while the plasma potential outside this volume is, essentially, unaffected. This model has later been supported by plasma potential measurements. [9] The ions that are accelerated are thus those that reach the boundary between the two plasma volumes during the time at which $U_{rev}$ is applied. The observation presented in the present work, that the portion of ions that are accelerated increases with $U_{rev}$ is an indication that the plasma volume close to the target have a weak electric potential driving ions through the plasma towards the main potential fall. Ions leaving the vicinity of the target during the HiPIMS pulse or after the reversed pulse, will not be accelerated and are the main constituent of the lower energy peak (and shoulder).

Regarding the slight decrease in deposition rate observed in this study, it is different from what has been observed for Cu-deposition by Wu *et al*. [11] who observe an increase in rate with $U_{rev}$, but similar to what we observe for Ti [10] and to what Velicu *et al*. [9] has reported for Cu. The reason for the different observation trends observed by different groups is unclear today, but we suspect that difference in magnetic field configuration as well as in pulse power arrangement are the cause for the deviating results.

A small effect on the apparent deposition rate loss can come from the densification of the film, see Fig. 4, that occurs when the film is subjected to bombardment due to atom insertion into the film subsurface region [24] and grain boundaries [25]. In the present work, low density grain boundaries are indeed observed in the case of $U_{rev} = 0$ and disappears as a reversed voltage is applied, see the HAADF-STEM images in Fig. 5.

The XRD studies of the deposited thin films show that the films are polycrystalline and there are no significant change in orientation texture between film grown with or without the reversed pulse. Previous works on HiPIMS-deposited TiN films show that it is possible to change the texture by applying a substrate bias [26] at a high current density [16]. The difference between these studies and the present experiment is that an elevated substrate temperature of 300 and 450 °C, respectively were used. In the present the substrates were not intentionally heated and the temperature is likely to be below 150 °C during the deposition process.

Several studies [26,27] on TiN films have shown that a change in texture is often accompanied by a change in the stress of the thin films such that for TiN (001) oriented grains are more susceptible to stress generation than (111) oriented grains since they are more open and survive intense ion bombardment [28] due to ion channeling effect. In the present work, regardless of no change in texture there is a significant change in hardness and in compressive stress when switching to the bipolar mode. Considering the densification of the grain boundaries of the growing film as the ion energy increases, an increase in the compressive stress can be expected. Finally we note that, as a consequence of densification and increased stress, the film hardness is improved significantly.

**Conclusions**

The effect of operating the HiPIMS process in bipolar mode is clearly demonstrated in the reactive sputter deposition of TiN thin films. An increased film density, hardness and compressive stress are the result of the application of the reversed voltage pulse following the HiPIMS pulse. IEDF measurements show that a large portion of the ions generated by the HiPIMS pulses are accelerated with the full applied potential $U_{rev}$. Varying $U_{rev}$ from 0 (conventional HiPIMS) to 150 V did not result in any change of the film texture but resulted in an increase in ion energies, in a densification of grain boundaries, in an increase of the film compressive stress from 2.1 to 4.7 GPa and in an increase in the hardness from 23.9 to 34 GPa.

**Acknowledgements**

This work has been supported by the Swedish Research Council (grant VR 621-2014-4882), the Swedish Government Strategic Research Area in Materials Science on Functional materials at Linköping University (Faculty Grant SFO-Mat-LiU No. 2009-00971) and the Beihang University Doctoral Student Short-Term Visiting Fund. We would also like to thank Dr. Fredrik Eriksson for the assistance during the XRR measurements and Ionautics AB for supplying the pulsing equipment which made the experiments possible.

**References**

[1] V. Kouznetsov, K. Macák, J.M. Schneider, U. Helmersson, I. Petrov, A novel pulsed magnetron sputter technique utilizing very high target power densities, Surf. Coat. Technol. 122 (1999) 290–293. doi:10.1016/S0257-8972(99)00292-3.

[2] V. Stranak, A.-P. Herrendorf, H. Wulff, S. Drache, M. Cada, Z. Hubicka, M. Tichy, R. Hippler, Deposition of rutile (TiO2) with preferred orientation by assisted high power impulse magnetron sputtering, Surf. Coat. Technol. 222 (2013) 112–117. doi:10.1016/j.surfcoat.2013.02.012.

[3] F. Cemin, G. Abadias, T. Minea, C. Furgeaud, F. Brisset, D. Solas, D. Lundin, Benefits of energetic ion bombardment for tailoring stress and microstructural evolution during growth of Cu thin films, Acta Mater. 141 (2017) 120–130. doi:10.1016/j.actamat.2017.09.007.

[4] J. Alami, P. Eklund, J.M. Andersson, M. Lattemann, E. Wallin, J. Bohlmark, P. Persson, U. Helmersson, Phase tailoring of Ta thin films by highly ionized pulsed magnetron sputtering, Thin Solid Films. 515 (2007) 3434–3438. doi:10.1016/j.tsf.2006.10.013.

[5] J. Bohlmark, J. Alami, C. Christou, A.P. Ehiasarian, U. Helmersson, Ionization of sputtered metals in high power pulsed magnetron sputtering, J. Vac. Sci. Technol. Vac. Surf. Films. 23 (2005) 18–22. doi:10.1116/1.1818135.

[6] D. Lundin, M. Čada, Z. Hubička, Ionization of sputtered Ti, Al, and C coupled with plasma characterization in HiPIMS, Plasma Sources Sci. Technol. 24 (2015) 035018. doi:10.1088/0963-0252/24/3/035018.

[7] M.M.S. Villamayor, J. Keraudy, T. Shimizu, R.P.B. Viloan, R. Boyd, D. Lundin, J.E. Greene, I. Petrov, U. Helmersson, Low temperature ($T_s / T_m < 0.1$) epitaxial growth of HfN/MgO(001) via reactive HiPIMS with metal-ion synchronized substrate bias, J. Vac. Sci. Technol. A. 36 (2018) 061511. doi:10.1116/1.5052702.

[8] G. Greczynski, I. Zhirkov, I. Petrov, J.E. Greene, J. Rosen, Control of the metal/gas ion ratio incident at the substrate plane during high-power impulse magnetron sputtering of transition metals in Ar, Thin Solid Films. 642 (2017) 36–40. doi:10.1016/j.tsf.2017.09.027.

[9] I.-L. Velicu, G.-T. Ianoş, C. Porosnicu, I. Mihăilă, I. Burducea, A. Velea, D. Cristea, D. Munteanu, V. Tiron, Energy-enhanced deposition of copper thin films by bipolar high power impulse magnetron sputtering, Surf. Coat. Technol. 359 (2019) 97–107. doi:10.1016/j.surfcoat.2018.12.079.

[10] J. Keraudy, R.P.B. Viloan, M.A. Raadu, N. Brenning, D. Lundin, U. Helmersson, Bipolar HiPIMS for tailoring ion energies in thin film deposition, Surf. Coat. Technol. 359 (2019) 433–437. doi:10.1016/j.surfcoat.2018.12.090.

[11] B. Wu, I. Haehnlein, I. Shchelkanov, J. McLain, D. Patel, J. Uhlig, B. Jurczyk, Y. Leng, D.N. Ruzic, Cu films prepared by bipolar pulsed high power impulse magnetron sputtering, Vacuum. 150 (2018) 216–221. doi:10.1016/j.vacuum.2018.01.011.

[12] N. Britun, M. Michiels, T. Godfroid, R. Snyders, Ion density evolution in a high-power sputtering discharge with bipolar pulsing, Appl. Phys. Lett. 112 (2018) 234103. doi:10.1063/1.5030697.

[13] A.J. Rosakis, R.P. Singh, Y. Tsuji, E. Kolawa, N.R.M. Jr, Full field measurements of curvature using coherent gradient sensing: application to thin film characterization, Thin Solid Films. (1998) 13. doi:https://doi.org/10.1016/S0040-6090(98)00432-5.

[14] T. Dieing, B.F. Usher, Radius of curvature in MBE grown heterostructures, in: COMMAD 2000 Proc. Conf. Optoelectron. Microelectron. Mater. Devices, IEEE, Bundoora, Vic., Australia, 2000: pp. 214–217. doi:10.1109/COMMAD.2000.1022930.

[15] G.C.A.M. Janssen, M.M. Abdalla, F. van Keulen, B.R. Pujada, B. van Venrooy, Celebrating the 100th anniversary of the Stoney equation for film stress: Developments from polycrystalline steel strips to single crystal silicon wafers, Thin Solid Films. 517 (2009) 1858–1867. doi:10.1016/j.tsf.2008.07.014.

[16] A.P. Ehiasarian, A. Vetushka, Y.A. Gonzalvo, G. Sáfrán, L. Székely, P.B. Barna, Influence of high power impulse magnetron sputtering plasma ionization on the microstructure of TiN thin films, J. Appl. Phys. 109 (2011) 104314. doi:10.1063/1.3579443.

[17] G. Greczynski, I. Zhirkov, I. Petrov, J.E. Greene, J. Rosen, Time evolution of ion fluxes incident at the substrate plane during reactive high-power impulse magnetron

sputtering of groups IVb and VIb transition metals in Ar/N $_2$, J. Vac. Sci. Technol. Vac. Surf. Films. 36 (2018) 020602. doi:10.1116/1.5016241.

[18]   A. Howie, Image Contrast And Localized Signal Selection Techniques, J. Microsc. 117 (1979) 11–23. doi:10.1111/j.1365-2818.1979.tb00228.x.

[19]   P. Poolcharuansin, J.W. Bradley, Short- and long-term plasma phenomena in a HiPIMS discharge, Plasma Sources Sci. Technol. 19 (2010) 025010. doi:10.1088/0963-0252/19/2/025010.

[20]   C. Maszl, W. Breilmann, J. Benedikt, A. von Keudell, Origin of the energetic ions at the substrate generated during high power pulsed magnetron sputtering of titanium, J. Phys. Appl. Phys. 47 (2014) 224002. doi:10.1088/0022-3727/47/22/224002.

[21]   A. Hecimovic, K. Burcalova, A.P. Ehiasarian, Origins of ion energy distribution function (IEDF) in high power impulse magnetron sputtering (HIPIMS) plasma discharge, J. Phys. Appl. Phys. 41 (2008) 095203. doi:10.1088/0022-3727/41/9/095203.

[22]   T. Nakano, N. Hirukawa, S. Saeki, S. Baba, Effects of target voltage during pulse-off period in pulsed magnetron sputtering on afterglow plasma and deposited film structure, Vacuum. 87 (2013) 109–113. doi:10.1016/j.vacuum.2012.03.010.

[23]   F. Avino, A. Sublet, M. Taborelli, Evidence of ion energy distribution shift in HiPIMS plasmas with positive pulse, Plasma Sources Sci. Technol. 28 (2019) 01LT03. doi:10.1088/1361-6595/aaf5c9.

[24]   M.M.M. Bilek, D.R. McKenzie, A comprehensive model of stress generation and relief processes in thin films deposited with energetic ions, Surf. Coat. Technol. 200 (2006) 4345–4354. doi:10.1016/j.surfcoat.2005.02.161.

[25]   D. Magnfält, G. Abadias, K. Sarakinos, Atom insertion into grain boundaries and stress generation in physically vapor deposited films, Appl. Phys. Lett. 103 (2013) 051910. doi:10.1063/1.4817669.

[26]   R. Machunze, A.P. Ehiasarian, F.D. Tichelaar, G.C.A.M. Janssen, Stress and texture in HIPIMS TiN thin films, Thin Solid Films. 518 (2009) 1561–1565. doi:10.1016/j.tsf.2009.09.069.

[27]   R. Machunze, G.C.A.M. Janssen, Stress and strain in titanium nitride thin films, Thin Solid Films. 517 (2009) 5888–5893. doi:10.1016/j.tsf.2009.04.020.

[28]   I. Petrov, L. Hultman, J. -E. Sundgren, J.E. Greene, Polycrystalline TiN films deposited by reactive bias magnetron sputtering: Effects of ion bombardment on resputtering rates, film composition, and microstructure, J. Vac. Sci. Technol. Vac. Surf. Films. 10 (1992) 265–272. doi:10.1116/1.578074.